\documentclass[mathleft]{an}
\usepackage{graphicx}
\usepackage{times,aas_macros}
\usepackage{multirow}
\usepackage{graphicx, subfigure}
\usepackage{natbib}
\bibpunct{(}{)}{;}{a}{}{,}
\begin{document}

% The following seven commands are intended for editorial usage and should be ignored by
% the author(s).
\Pagespan{1}{}% Document's page range. 
% If second parameter is left empty, the last page is computed automatically.
\Yearpublication{}%
\Yearsubmission{}%
\Month{}%   
\Volume{}%  
\Issue{}% 
% \DOI{This.is/not.aDOI}% 

\title{Red Giant Oscillations: Stellar Models and Mode Frequency Calculations}

\author{A. Jendreieck\inst{1}\fnmsep\thanks{Corresponding author:
  \email{ajendreieck@mpa-garching.mpg.de}\newline}
%Example 
%for footnote, note the usage of the \texttt{fnmsep}
%command as separator between institute number and footnote mark} 
\and  A. Weiss\inst{1} \and V. Silva Aguirre\inst{2} \and J. Christensen-Dalsgaard\inst{2} \and R. Handberg\inst{2} \and G. Ruchti\inst{1}
\and  C. Jiang\inst{2} \and A. Thygesen\inst{3}
}
\titlerunning{Red Giant Oscillations: Stellar Models and Mode Frequency Calculations}
\authorrunning{A. Jendreieck et al.}
\institute{
Max Planck Institute for Astrophysics, Karl-Schwarzschild-Str. 1, 85748, Garching bei M\"unchen, Germany 
\and 
Stellar Astrophysics Centre, Department of Physics and Astronomy, Aarhus University, Ny Munkegade 120, \\
DK-8000 Aarhus C, Denmark
\and
Zentrum f\"{u}r Astronomie der Universit\"{a}t Heidelberg, Landessternwarte, Heidelberg, Germany
}

\received{22 August 2012}
\accepted{18 October 2012}
\publonline{?}


\abstract{%
  We present preliminary results on modelling KIC 7693833, the so far most metal-poor red-giant star observed by {\it Kepler}. From time series spanning several months, global oscillation parameters and individual frequencies were obtained and compared to theoretical calculations. Evolution models are calculated taking into account spectroscopic and asteroseismic constraints. The oscillation frequencies of the models were computed and compared to the {\it Kepler} data. In the range of mass computed, there is no preferred model, giving an uncertainty of about 30 K in $T_{\mathrm{eff}}$, 0.02 dex in $\log g$, $0.7 \,R_{\mathrm{\odot}}$ in radius and of about 2.5 Gyr in age.}

\maketitle

\section{Introduction}
Red giants have a broad spectrum of acoustic oscillation modes which are excited by an extended convective envelope \citep[e.g.][]{montalban2010,mosser2011}. Also, as the density in the helium core is quite large, g-modes with frequencies close to those of the acoustic modes can interact with the latter, bringing on the presence of so-called mixed modes \citep{dimauro2011}. Those modes with a more p-mode character, propagating with smaller inertia, can reach the surface where they are observed. As they have g-mode character in the centre, they provide important information about the deep interior of red giants. The high sensitivity of the {\it CoRoT} and {\it Kepler} space missions made it possible to observe non-radial oscillations in a large sample of red giants (De Ridder et al., 2009) and the observation of such mixed modes \citep[e.g.][]{dimauro2011}. The study of evolved stars is important as it leads to better constraints on stellar evolution models, since the uncertainties in stellar structure properties accumulate with age \citep{deridder2009}. Processes not well understood like convective overshooting, rotational mixing and diffusion during the hydrogen-burning phase determines the mass of the helium core in the giant phase and also the age of the star \citep{aerts2008}.

\section{Observations}
We have analyzed the red giant KIC 7693833 observed by {\it Kepler} during the Q1-Q10. The frequency spectrum spans a range from $20\,\mu$Hz to $50\,\mu$Hz with 19 modes identified as $l=0,1,2$ with a $\nu_{\mathrm{max}}=32.2\,\mu$Hz and $\Delta \nu =4.06\,\mu$Hz. The target KIC 7693833 is a metal-poor star classified ($\left[\mathrm{Fe/H}\right]=-2.2$ dex) as ascending the red-giant branch according to the classification given by \citet{bedding2011}. Spectroscopic analysis was performed by \citet{thygensen2012} and a $T_{\mathrm{eff}}$ of about 4800 K was determined. We re-analyzed the spectra considering NLTE effects and found a $T_{\mathrm{eff}}$ of about 5100 K. We also found indications of a constant $\alpha$-enhancement of 0.2 dex. From the corrected {\it Kepler} lightcurves we extracted the global seismic parameters $\Delta \nu$ and $\nu_{\mathrm{max}}$, as well as individual frequencies for modes of degree $l=0,1,2$. The basic stellar parameters are summarized in Table \ref{par}.

\begin{table}
\centering
\caption{Basic parameters of KIC 7693833 determined and used in the present analysis.}
\label{par}
\begin{tabular}{cc}\hline \hline
\multicolumn{2}{c}{Basic Parameters of KIC 7693833} \\ 
\hline
$T_\mathrm{eff}$ & $5119\pm 140$ K \\
log g & $2.4\pm0.2$ dex\\
$[$Fe/H$]$ & $-2.20\pm 0.1$ dex \\
$\alpha$-enhancement & $0.2$ dex \\
$\nu_\mathrm{max}$ & $32.2\pm 2.0 \, \mu$Hz \\
$\Delta \nu$ & $4.06\pm 0.2 \, \mu$Hz \\
\hline \hline
\end{tabular}
\end{table}

\section{Models}
Making use of the observed constraints on the asteroseismic parameters and the effective temperature, we first estimated a preliminary mass using the asteroseismic scaling relation \citep{brown1991,kjeldsen1995}:
\begin{equation}
\frac{M}{M_\mathrm{\odot}}=\left(\frac{\Delta\nu}{\Delta\nu_\mathrm{\odot}}\right)^{-4}\left(\frac{\nu_\mathrm{{max}}}{\nu_\mathrm{{max,\odot}}}\right)^{3} \left(\frac{T_\mathrm{{eff}}}{T_\mathrm{{eff,\odot}}}\right) ^{3/2},
\end{equation}
resulting in a mass of $1.17$ $M_\odot$. With this information, a grid of models was calculated for masses from $1.00-1.20$ $M_\odot$ in steps of $0.01$ $M_\odot$ with the GARSTEC evolutionary code \citep{weiss2008}.

All the models were computed with the same input parameters. OPAL 2005 \citep{rogers2002} equation of state were used together with NACRE \citep{angulo1999} reaction rates. The metallicity was calculated considering the $\alpha$-enhancement according to Salaris et al. (1997), giving a $Z=1.6 \times 10^{-4}$. We used the He primordial content $Y=0.248$ derived by \citet{steigman2010}. Convection is treated according to the mixing-length theory (MLT) \citep{bohmvitense1958} and with the parameter $\alpha_{\mathrm{MLT}}=1.744$ calibrated for the Sun.

The oscillation frequencies were calculated with the ADIPLS package \citep{dalsgaard2008} for modes of angular degree $l=0,1$ and 2 for all models that satisfied $\Delta \nu$ observed according to the scaling relation:
\begin{equation}
\frac{\Delta\nu}{\Delta\nu_\mathrm{{\odot}}}=\left(\frac{M}{M_\mathrm{\odot}}\right)^{1/2}\left(\frac{L}{L_\mathrm{\odot}}\right)^{-0.75}\left(\frac{T_\mathrm{{eff}}}{T_\mathrm{{eff,\odot}}}\right)^3.
\end{equation}
Afterwards, a thinner time step was set to find the models that matched the lowest $l=0$ mode observed within the error bars.

\section{Results to Date}
\begin{figure}
\includegraphics[scale=0.50]{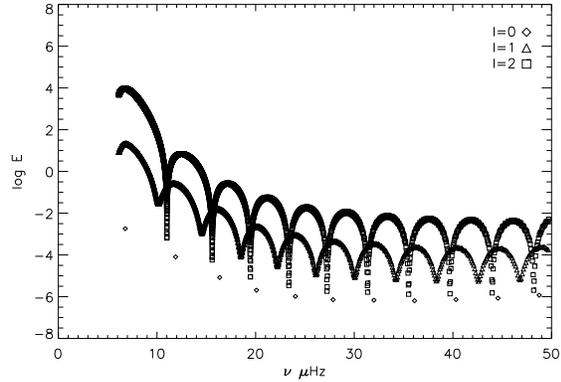}
\caption{Mode inertia as function of frequency for modes $l=0$ (diamonds), $l=1$ (triangles) and $l=2$ (squares). The modes with lowest inertia are the ones with higher amplitudes at the surface and hence are easier to observe.}
\label{inertia}
\end{figure}

We compared the calculated and observed frequencies. As the calculated frequency spectrum is heavily populated, only the frequencies with minimum mode inertia were chosen to be compared because these modes have the highest amplitudes at the surface and hence are the most probable to be observed \citep[e.g.][]{dalsgaard1995,dupret2009}. Figure \ref{inertia} shows mode inertia plotted versus frequency for a model with 1 $M_{\mathrm{\odot}}$. Radial modes are represented as diamonds, $l=1$ modes as triangles and $l=2$ modes as squares. 

\begin{figure*}
\subfigure{\includegraphics[scale=0.50]{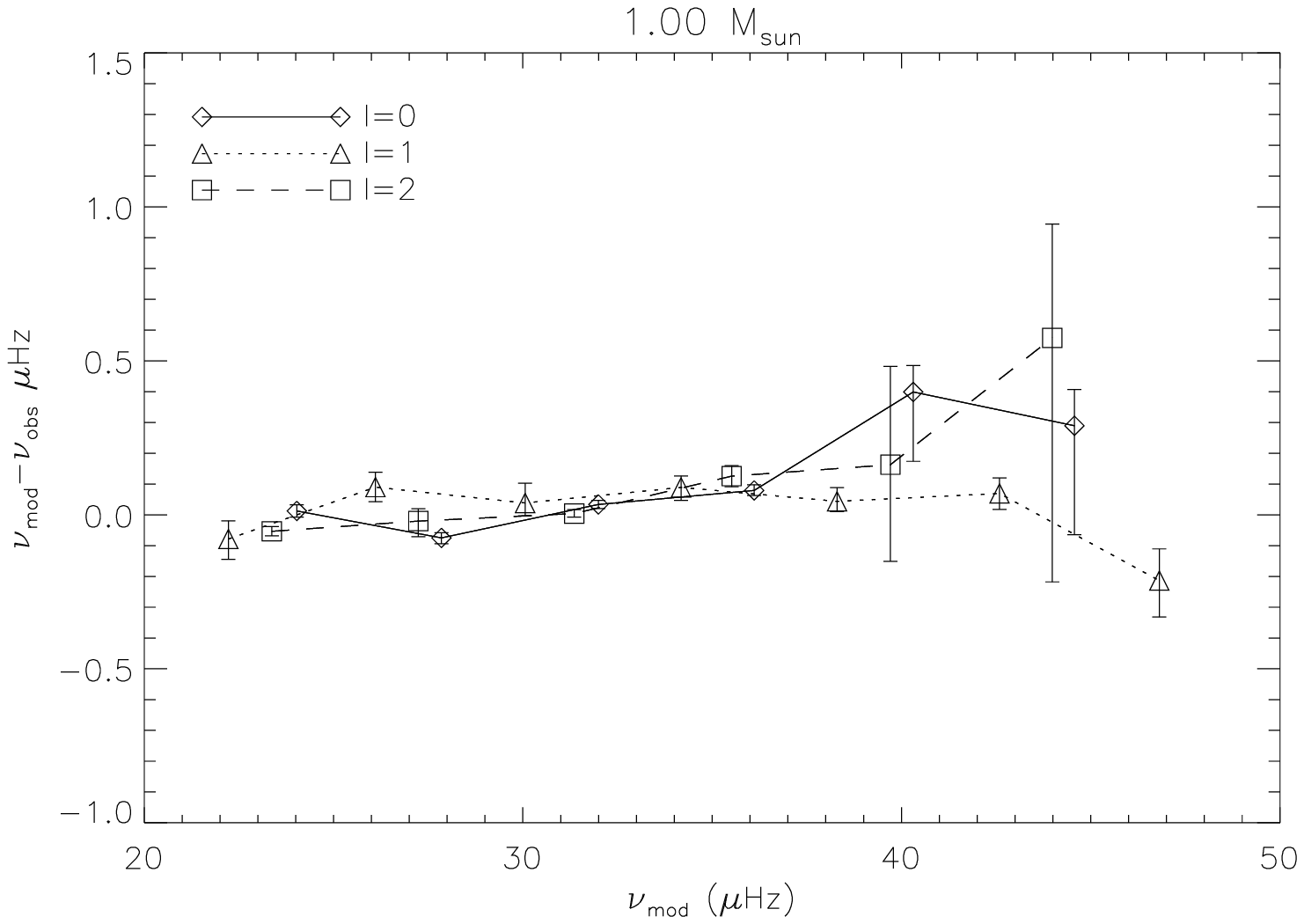}}
\subfigure{\includegraphics[scale=0.50]{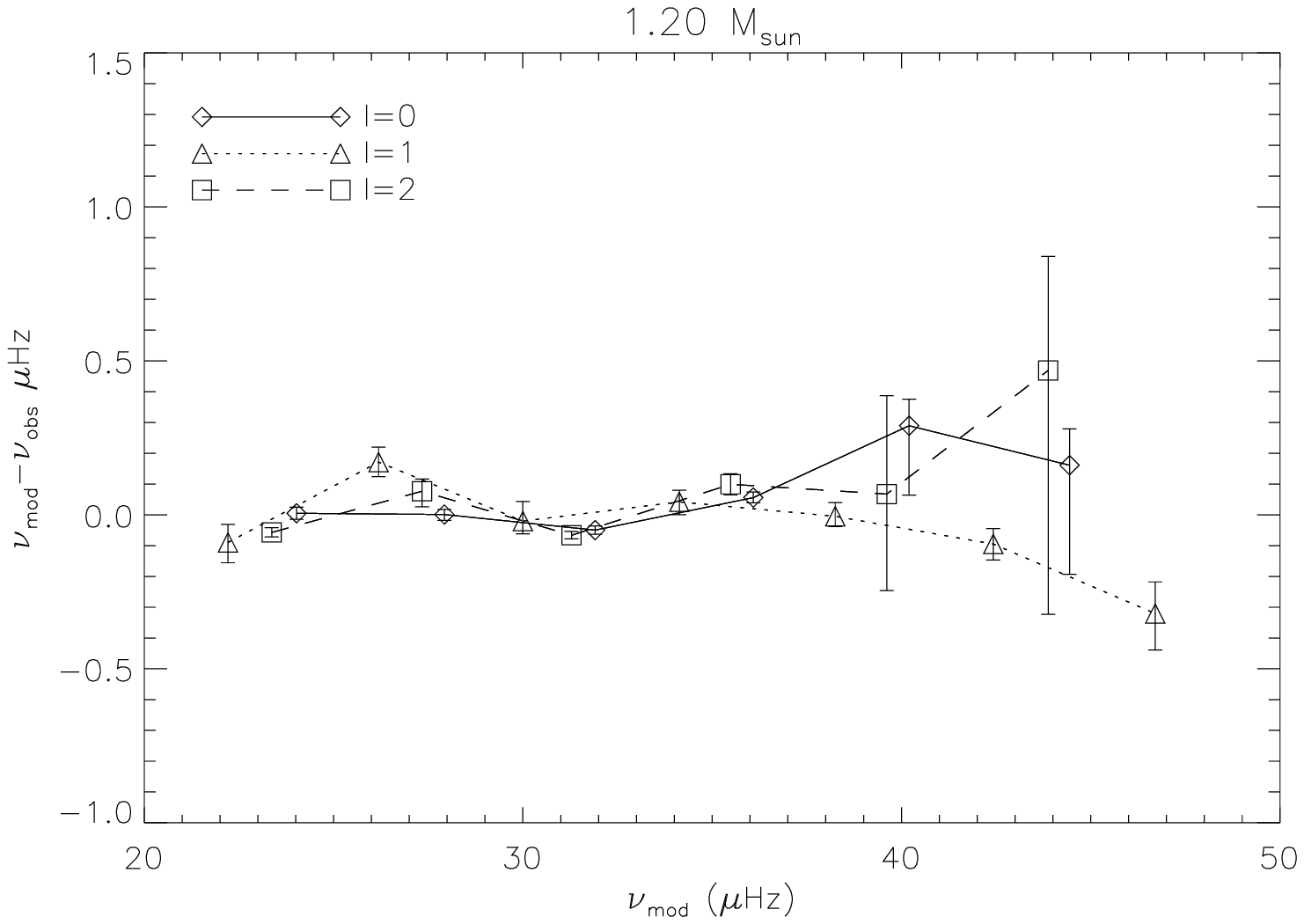}}
\caption{Frequency difference between model and observational values for models with 1.00 (left) and 1.20 (right) $\mathrm{M}_\odot$. Both models agree within a range of $1.0 \,\mu$Hz with the data. There is no preferred model.}
\label{dif}
\end{figure*}

\begin{figure*}
\subfigure{\includegraphics[scale=0.50]{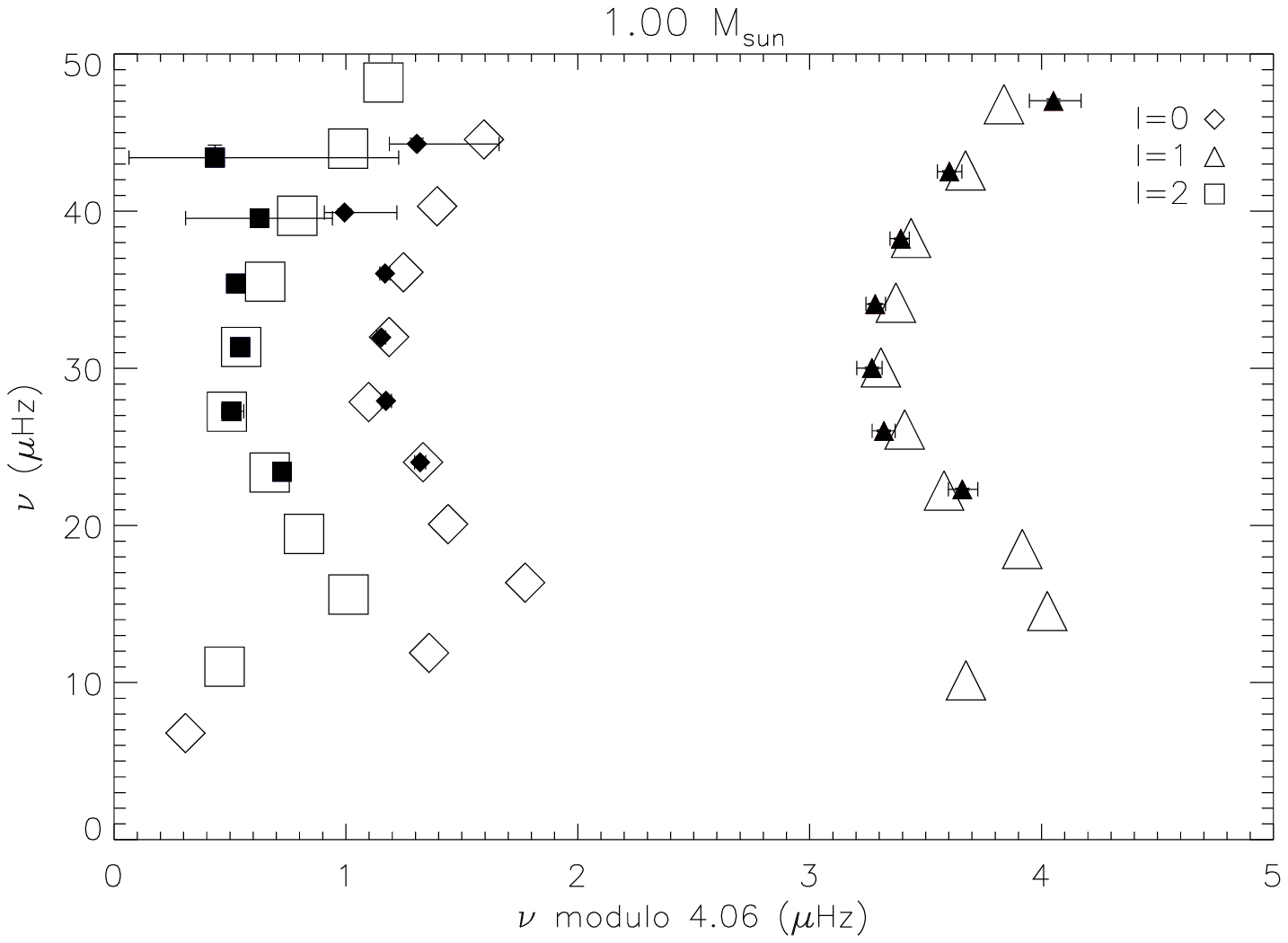}}
\subfigure{\includegraphics[scale=0.50]{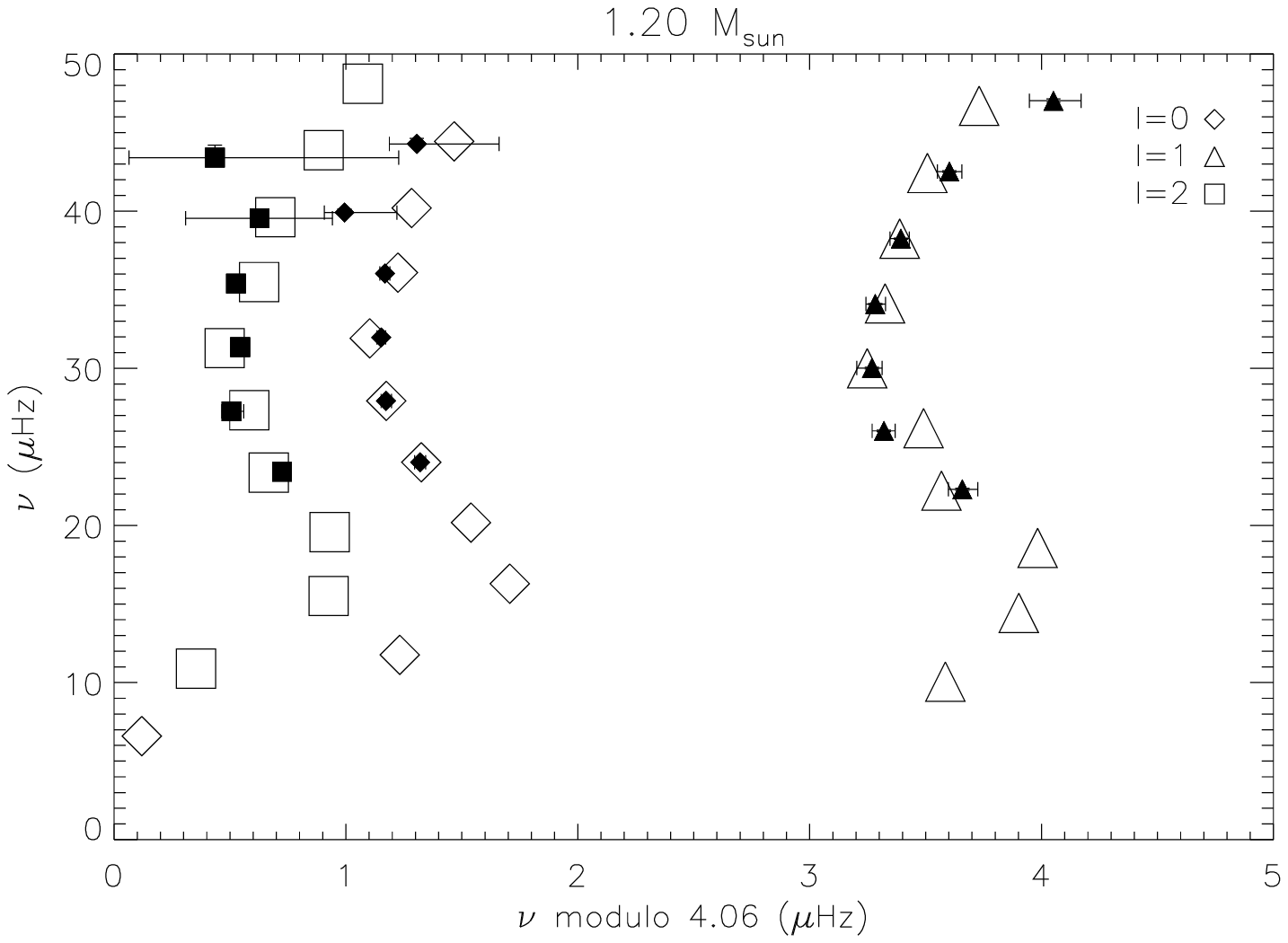}}
\caption{\'Echelle diagram for the models with 1.00 M$_{\mathrm{\odot}}$ (left) and 1.20 M$_{\mathrm{\odot}}$ (right). Open symbols show the frequencies computed for the models and filled symbols show the observed ones. Diamonds are for $l=0$ modes, triangles for $l=1$ modes and squares for $l=2$.}
\label{echelle}
\end{figure*}

Figure \ref{dif} shows the difference between calculated and observed frequencies, for some of the models with masses $1.0$ and $1.2$ M$_\mathrm{\odot}$. The error bars are representing the errors from the observational data. The frequency differences are smaller than $1.0\,\mu$Hz for all models considered up to now, having no preferred model. Figure \ref{echelle} shows the \'echelle diagrams for the same models. Open symbols are the frequencies computed from models and filled circles are representing the observational data. Both models show good agreement with data, showing again no preferred model.

In the range of masses investigated by now, the discrepancies in the determination of the basic parameters are showed in Table \ref{disc}.

\begin{table}
\centering
\caption{Differences in the basic parameters for the models with 1.00 M$_{\mathrm{\odot}}$ and 1.20 M$_{\mathrm{\odot}}$}
\label{disc}
\begin{tabular}{ccc}\hline \hline
Mass & 1.00 M$_{\mathrm{\odot}}$ & 1.20 M$_{\mathrm{\odot}}$ \\ 
\hline
$T_\mathrm{eff}$ & $5159$ K & $5190$ K \\
$\log g$ & $2.42$ dex & $2.44$ dex\\\
$R/R_\odot$ & $10.2$ & $10.9$ \\
Age & $5.7$ Gyr & 3.2 Gyr  \\
\hline \hline
\end{tabular}
\end{table}

We also tried to investigate the asymptotic period spacing of the models to see if it would be a good way to disentangle between the mass. However the difference between 1.00 and 1.20 M$_{\mathrm{\odot}}$ are of a few seconds, both about 60\,s. \citet{bedding2011} found a value of observed period spacing  of 80\,s giving an even larger asymptotic value. The discrepancies in period spacing might be solved with higher mass and is being investigated.

\section{Conclusions and Perspectives}
KIC 7693833 is the so far most metal-poor red giant and it was observed by {\it Kepler} during Q1-Q10. From its frequency spectrum, 19 modes of oscillation were identified as $l=0,1,2$ with a $\nu_{\mathrm{max}}=32.2 \,\mu$Hz and $\Delta \nu =  4.06 \,\mu$Hz.

Models within 1.00-1.20 M$_{\mathrm{\odot}}$ were computed to match the observational constraints using the scaling relations as starting point. The frequencies computed for all models agree within $1.0 \,\mu$Hz, having no preferred model. This gives an uncertainty of about 30 K in $T_{\mathrm{eff}}$, 0.02 dex in $\log g$, 0.7 $R_{\mathrm{\odot}}$ in radius, corresponding to 7\%, and of about 2.5 Gyr in age. The period spacing of the models computed are smaller than the observed one and the possibility to resolve the discrepancies by possible higher masses is currently investigated. This would make the star even younger and quite peculiar for a metal-poor star.

For a more detailed modeling, it is important to investigate different physics inputs and see how the oscillation frequencies change. This way, we might find a way to distinguish between models of different masses and different physics. Currently, we still have no good diagnosis to disentangle models of red giants.

\acknowledgements
Funding for the {\it Kepler} Discovery mission is provided by NASA's Science Mission Directorate.
Funding for the Stellar Astrophysics Centre is provided by The Danish National Research Foundation. The research is supported by the ASTERISK project (ASTERoseismic Investigations with SONG and Kepler) funded by the European Research Council (Grant agreement no.: 267864).

%\newpage%%%%%%%%%%%%%%%%%%%%%%%%%%%%%%%%%%%%%%%%%%%%%%%%%%%%%%

\bibliographystyle{mn}
\bibliography{bibliografia1}

\end{document}